\documentclass{aastex631}

\usepackage{graphicx}
\usepackage{color}
\usepackage{epsfig}
\usepackage{epsf}
\usepackage{epstopdf}
\usepackage{subfigure}
\usepackage{upgreek}
\usepackage{amssymb}
\usepackage{amsmath}
\usepackage{natbib}
\usepackage{graphicx}
\usepackage{color}
\usepackage{tabularx}
\usepackage{rotating}
\usepackage{multirow}
\usepackage{subfigure}
\usepackage{ulem}
\usepackage{longtable}

\begin{document}

\title{Period--Luminosity Relations for Double-mode $\delta$ Sct Stars}

\author[0000-0002-0786-7307]{Qi Jia}
\affiliation{School of Physics and Astronomy, China West Normal University, Nanchong 637009, China}
\affiliation{CAS Key Laboratory of Optical Astronomy, National Astronomical Observatories, Chinese Academy of Sciences, Beijing 100101,  China}

\author[0000-0001-7084-0484]{Xiaodian Chen}
\affiliation{CAS Key Laboratory of Optical Astronomy, National Astronomical Observatories, Chinese Academy of Sciences, Beijing 100101,  China}
\affiliation{School of Physics and Astronomy, China West Normal University, Nanchong 637009, China}
\affiliation{School of Astronomy and Space Science, University of the Chinese Academy of Sciences, Beijing, 100049, China}
\affiliation{Institute for Frontiers in Astronomy and Astrophysics, Beijing Normal University,  Beijing 102206, China}

\author[0000-0003-4489-9794]{Shu Wang}
\affiliation{CAS Key Laboratory of Optical Astronomy, National Astronomical Observatories, Chinese Academy of Sciences, Beijing 100101,  China}
\affiliation{School of Physics and Astronomy, China West Normal University, Nanchong 637009, China}

\author{Licai Deng}
\affiliation{CAS Key Laboratory of Optical Astronomy, National Astronomical Observatories, Chinese Academy of Sciences, Beijing 100101,  China}
\affiliation{School of Physics and Astronomy, China West Normal University, Nanchong 637009, China}
\affiliation{School of Astronomy and Space Science, University of the Chinese Academy of Sciences, Beijing, 100049, China}

\author{Jianxing Zhang}
\affiliation{CAS Key Laboratory of Optical Astronomy, National Astronomical Observatories, Chinese Academy of Sciences, Beijing 100101,  China}

\affiliation{School of Astronomy and Space Science, University of the Chinese Academy of Sciences, Beijing, 100049, China}

\author{Qingquan Jiang}
\affiliation{School of Physics and Astronomy, China West Normal University, Nanchong 637009, China}

\correspondingauthor{Xiaodian Chen}
\email{chenxiaodian@nao.cas.cn}



\begin{abstract}
Previous studies of the Period--Luminosity relations (PLRs) of Delta Scuti ($\delta$ Sct) stars have focused on those with a single pulsation mode. However, for $\delta$ Sct stars with many different pulsation modes, classifying a single mode is difficult. In this study, an all-sky dataset is constructed using double-mode $\delta$ Sct stars from ZTF and OGLE, and is used to determine F-mode and 1O-mode PLRs for eight single bands and six Wesenheit bands. In the $W1$ band, the PLR dispersion is about 0.171 mag and the total zero point error is 1\%. Our results show that to accurately classify the 1O modes of $\delta$ Sct stars requires authentication based on multiple modes. Classification based on amplitude alone leads to impure 1O-mode $\delta$ Sct stars and significant deviations in the PLRs. We compare the PLRs of the different sequences in the Petersen diagram and find that they are consistent after a strict criterion filtering, suggesting that their evolutionary state is similar. In addition, we find a weak period--metallicity relation for double-mode $\delta$ Sct stars, unlike double-mode RR Lyrae stars. As distance tracers, large-amplitude F-mode $\delta$ Sct and double-mode $\delta$ Sct stars are the most suitable of the $\delta$ Sct family.
\end{abstract}


\section{introduction}
The period--luminosity relations (PLR) are essential for cosmic distance measurement and have been evolving since Leavitt's discovery in the 1920s \citep{1912HarCi.173....1L}. Pulsating variable stars exhibit periodic brightness variations due to internal pulsations, are reliable standard candles for measuring cosmic distances and understanding the Milky Way structure \citep{2019NatAs...3..320C}. The study of PLRs in classical Cepheids, RR Lyrae stars, and the luminosity of Type Ia supernovae has contributed significantly to our knowledge of the expansion of the Universe and the nature of dark energy \citep{2001ApJ...553...47F, 2019ApJ...876...85R, 2004ApJS..154..633C, 2015MNRAS.454.1509M, 2016AJ....151..118V, 2017ApJ...850..137M, 2018ApJ...852...78W, 2021CQGra..38o3001D, 2022ApJ...934L...7R}.

Despite their lower luminosity, $\delta$ Sct stars are valuable for distance measurements in the  due to their short periods and ease of observation \citep{2000ASPC..210....3B, 2000PASP..112.1096M, 2000ASPC..210..373M, 2001aa366..178R, 2004aa414L..17D, 2011aa534A.125U, 2017ampm.book.....B}. These stars typically exhibit radial and non-radial pressure modes influenced by the $\kappa$ mechanism \citep{2000ASPC..210....3B}, resulting in complex periodic variations. Many $\delta$ Sct stars show multiple modes, and the identification of their modes is important \citep{1973aa.23..221B, 1979PASP...91....5B, 1996aa312..463P, 1999aa341..151B, 1999MNRAS.302..349B, 2000ASPC..210....3B, 2000aa..144..469R, 2005aa438..653D}. The Petersen diagram helps to distinguish between different radial pulsation periods for multimode $\delta$ Sct stars \citep{1973aa.27...89P}.

The development of advanced optical telescopes, including ground-based facilities like the Zwicky Transient Facility \citep[ZTF,][]{2019PASP..131a8002B, 2020ApJS..249...18C} and the Optical Gravitational Lensing Experiment \citep[OGLE,][]{2015AcA....65....1U}, as well as space missions like Kepler \citep{2010Sci...327..977B, 2010ApJ...725.1226S} and Gaia \citep{2016aa595A...1G}, has enriched our understanding of the PLR of periodic variable stars. Spectroscopic surveys such as \citep[LAMOST,][]{2012RAA....12.1197C} and the Sloan Digital Sky Survey \citep[SDSS,][]{2000AJ....120.1579Y} have also provided extensive metallicity data, facilitating studies of the period--metallicity--luminosity relations. Previous studies of $\delta$ Sct stars have determined the PLRs of their fundamental (F) modes or first overtone (1O) modes \citep{1975ApJ...200..343B, 1991IBVS.3562....1K, 1997PASP..109.1221M, 2008ApJ...685..947P, 2010AcA....60....1P, 2010AJ....140..328G, 2011AJ....142..110M, 2012MNRAS.419..342C, 2019MNRAS.486.4348Z, J20, 2021PASP..133h4201P, 2022MNRAS.516.2080B, 2022ApJ...940L..25M, 2023aa...674A..36G, 2023arXiv230915147S, 2023AJ....165..190N}. These works have mainly investigated single-period $\delta$ Sct stars, or other periods of $\delta$ Sct stars have not been identified. Recently, \cite{2023NatAs...7.1081C} reported a robust linear relationship in the period--period ratio--metallicity relation for double-mode RR Lyrae stars (RRds), suggesting a reliable metallicity-independent distance indicator.

In this work, we combine the established multimode $\delta$ Sct sample in the northern sky \citep{2024ApJS..273....7J} and the multimode $\delta$ Sct sample from the OGLE catalog \citep{2020AcA....70..101S, 2021AcA....71..189S, 2022AcA....72..245S, 2023AcA....73..105S} with Gaia parallaxes and metallicity data from LAMOST. Our aim is to study the PLR of double-mode $\delta$ Sct stars and to investigate the effect of metallicity on both periods. The paper is organized as follows: In Section \ref{sample}, we describe in detail the double-mode $\delta$ Sct sample used. Section \ref{method} describes the data selection criteria and the fit method. Our fitting results are shown in Section \ref{result}. In Section \ref{dis}, we compare the PLR with previous studies and analyze the correlation between the periods and metallicities for double-mode $\delta$ Sct stars. We conclude our summary in Section \ref{conclusion}.

\begin{figure*}
	\begin{center}
\includegraphics[width=1.0\linewidth]{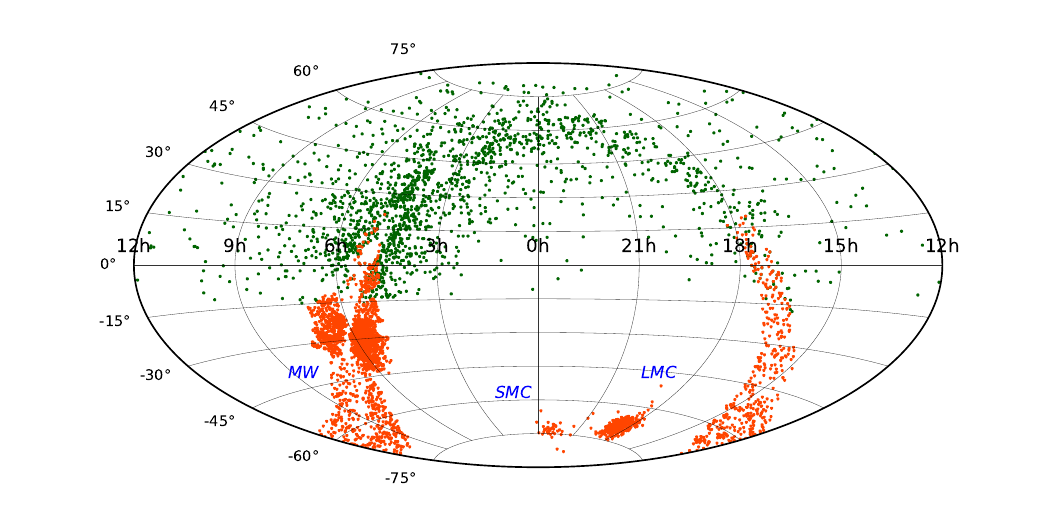}
		\caption{All-sky distribution of double-mode $\delta$ Sct stars in equatorial coordinates. Green dots indicate samples from ZTF and orange dots indicate samples from OGLE.}
		\label{sky}
	\end{center}
\end{figure*}

\begin{figure*}
	\begin{center}
\includegraphics[width=1.0\linewidth]{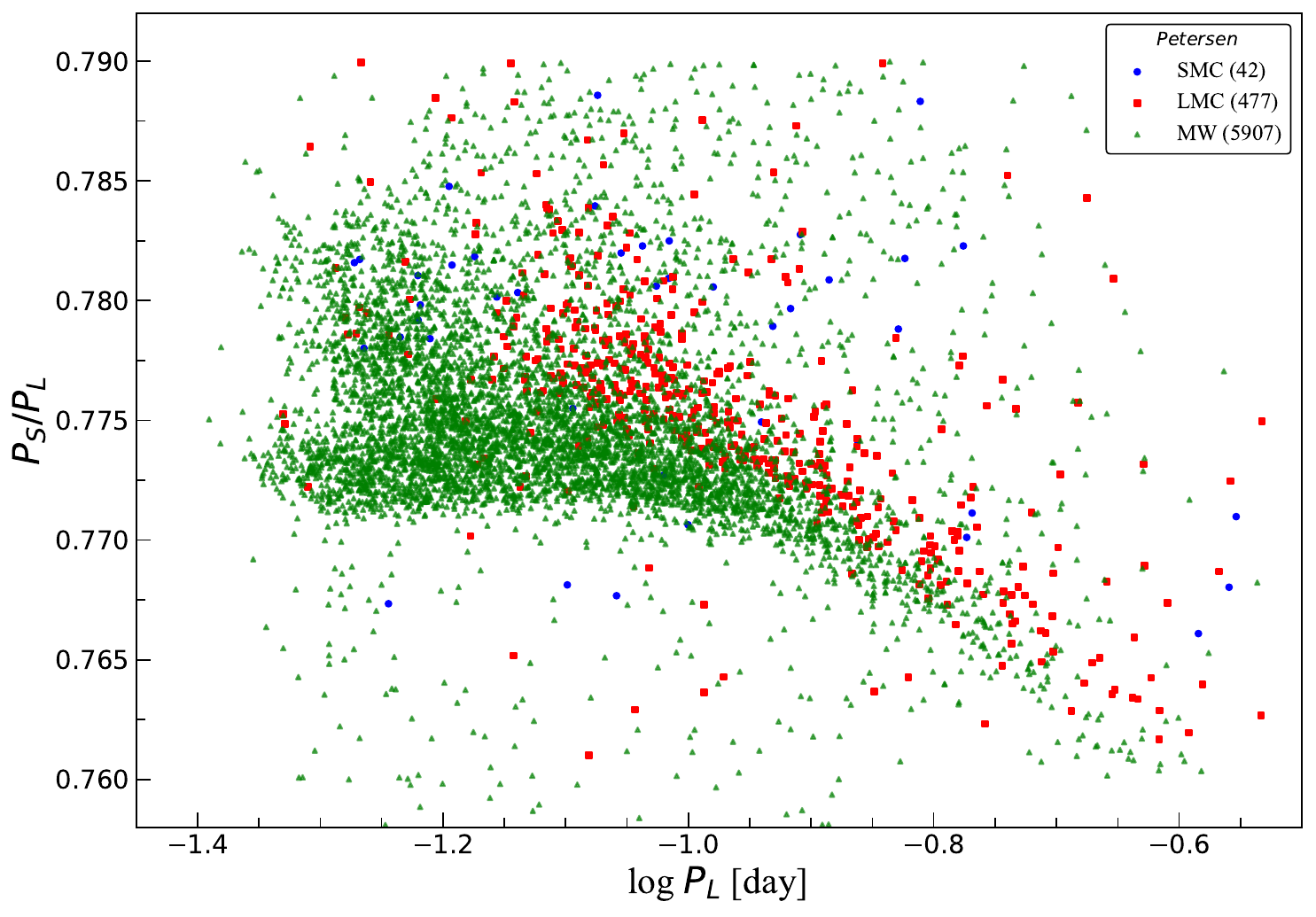}
		\caption{Petersen diagram of double-mode $\delta$ Sct stars. The period ratio (y axis) is the short period ($P_S$) divided by the long period ($P_L$). Blue circles denote the positions of $\delta$ Sct located in the SMC, red squares represent those in the LMC, green triangles indicate $\delta$ Sct in the MW, and gray points mark the positions of $\delta$ Sct not associated with the SMC, LMC, or MW.}
		\label{petersenlc}
	\end{center}
\end{figure*}

\section{Sample construction}
\label{sample}
\subsection{Sample Selection}
We collected 1838 $\delta$ Sct stars with F and 1O pulsation modes identified in the ZTF and added 4588 OGLE double-mode $\delta$ Sct stars in the southern sky. We obtained an all-sky sample of 6,426 double-mode $\delta$ Sct stars with their spatial distribution shown in Fig. \ref{sky}, where the green dots represent the ZTF sample and the orange dots represent OGLE sample. In the OGLE sample, there are 4069, 477, and 42 sources from the Milky Way, Large Magellanic Cloud (LMC), and Small Magellanic Cloud (SMC), respectively. Note that the LMC and SMC samples have a small amount of foreground $\delta$ Sct stars. We mapped all the sampled data on the Petersen diagram, as illustrated in Figure \ref{petersenlc}. The blue circles represent double-mode $\delta$ Sct stars in the SMC, red squares denote their counterparts in the LMC and green triangles indicate double-mode $\delta$ Sct stars within the Milky Way. The SMC contains relatively few double-mode $\delta$ Sct stars, while the primary difference between the LMC and Milky Way distributions is the lack of short-period double-mode $\delta$ Sct stars in the LMC. This deficiency is likely due to their fainter magnitudes, falling below OGLE’s detection limit for variable stars. The majority of the sample is concentrated within the Milky Way. In this study, we focus on analyzing the properties of Milky Way double-mode $\delta$ Sct stars using Gaia parallaxes to better characterize their intrinsic features.

The F-mode periods in our sample of double-mode $\delta$ Sct stars range from 0.04068 to 0.29347 days (approximately 0.98 to 7.04 hours), while the 1O-mode periods span 0.03153 to 0.22743 days (approximately 0.76 to 5.46 hours). Given the intrinsically short periods of $\delta$ Sct stars, concerns arise regarding whether the exposure times and cadences of large-scale surveys such as ZTF and OGLE could introduce challenges in period determination. Although the sampling intervals of both surveys may exceed the pulsation periods of $\delta$ Sct stars, their high cadence and long-term coverage generally enable robust period recovery. ZTF DR20, in particular, has a temporal resolution and survey duration that allow for the detection of variable stars with periods as short as $\sim0.02$ days. To assess the reliability of our period measurements, we compared them with the TESS variable star catalog \citep{2025ApJS..276...57G} and found good agreement of two periods for double-mode $\delta$ Sct stars. Furthermore, we have performed a detailed consistency analysis of the periods derived from OGLE and ZTF for common sources \citep{2024ApJS..273....7J}, confirming that the majority of the measured periods are highly consistent and reliable.

\subsection{Petersen Diagram}
\label{petersen detail}
\begin{figure*}
	\begin{center}
\includegraphics[width=1.0\linewidth]{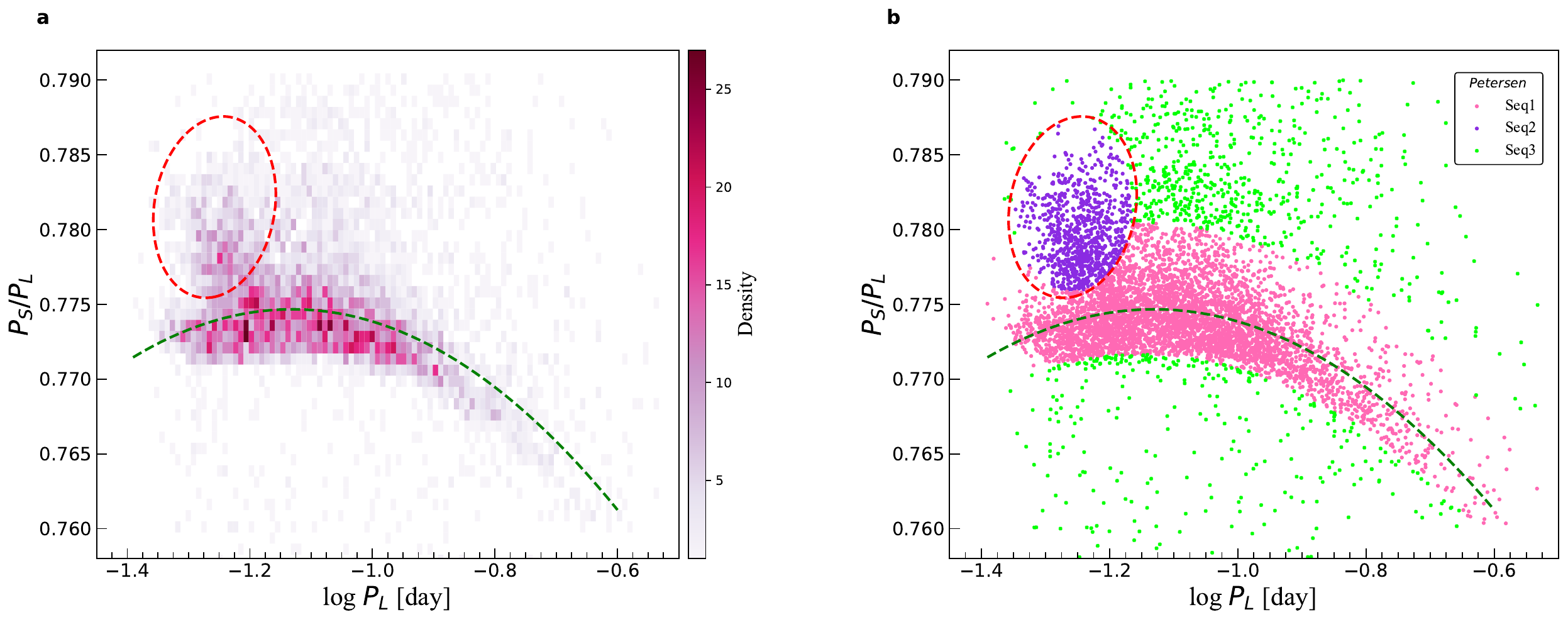}
		\caption{Petersen diagram of double-mode $\delta$ Sct stars. The period ratio (y axis) is the short period ($P_S$) divided by the long period ($P_L$). Panel a shows the density heat map of the Petersen diagram. Panel b shows the selected three sequences from the distribution, pink dots: Seq 1, purple dots: Seq 2, green dots: Seq 3.}
		\label{petersen}
	\end{center}
\end{figure*}

We draw the Petersen diagrams in Fig. \ref{petersen} based on 6426 $\delta$ Sct stars with F and 1O pulsation modes and find that the Petersen diagrams demonstrated three different distributions, a primary sequence (Seq 1), a clump (Seq 2) and other more dispersed points (Seq 3). Seq 1 is similar in style to the distribution of the double-mode RR Lyrae and classical Cepheids \citep{2023NatAs...7.1081C,2016MNRAS.460.2077K}, which are the typical double-mode variable stars. Their period ratios can be predicted by the pulsation models. Seq 3 exhibits a more scattered distribution around Seq 1, likely due to the influence of higher-order pulsation modes on their secondary periods. Seq 2 represents a distinct feature among double-mode $\delta$ Sct stars. However, for these two sequences, the correlation between their two periods remains uncertain. This study aims to explore potential evolutionary differences among these sequences by comparing their PLRs, using luminosity as a proxy.

To separate the three sequences, we first use an ellipse to separate the Seq 2. Based on the distribution of the density heat map, the selection criterion is shown in Equation \ref{cir}. The ellipse is centered at $\log{P}_{L} = -1.2576$ and $P_{S}/ P_{L} = 0.7815$, and we set the ellipse angle to 0.5$^\circ$ in the clockwise direction. For the remaining $\delta$ Sct stars, we performed a \texttt{ROBUST} regression using Scikit-learn's library\footnote{\url{https://scikit-learn.org/stable/index.html}} and identified $\delta$ Sct stars as the Seq 1 based on the regression dispersion. Our \texttt{ROBUST} regression approach incorporated a second-degree polynomial transformation (\texttt{PolynomialFeatures(degree=2)}) combined with the RANSAC algorithm (\texttt{RANSACRegressor}) to reduce the impact of outliers. The RANSAC regressor was configured with a \texttt{residual\_threshold} of 0.0044, determined through preliminary data analysis, and a fixed \texttt{random\_state} of 1 to ensure reproducibility. The model was trained using $\log P_L$ as the input and $P_{S} /P_{L}$ as the target variable. Performance evaluation based on the mean squared error (MSE) yielded a robust regression error of 0.00003. The resulting polynomial equation is given in Equation \ref{line}. We divided the remaining $\delta$ Sct stars into Seq 3. We tested different selection criteria for Seq 1 and found that the PLR zero points remain consistent across various selection thresholds.

\begin{equation}
((\log{P} _{L} +1.2576)/0.2)^{2} +((P_{S}/ P_{L}-0.7815)/0.012)^{2} \le 0.22    \label{cir}
\end{equation}

\begin{equation}
P_{S} /P_{L} =0.1234 + 2.5678\log{P_L}  -1.2345  (\log P_L)^2    \label{line}
\end{equation}

In Fig. \ref{petersen}b, Seq 1, 2, and 3 are shown with pink, purple, and green dots, corresponding to the number of double-mode $\delta$ Sct stars of 4396, 872, and 1020, respectively. To ensure that the analysis is performed on a purer sample of double-mode $\delta$ Sct stars, we use Seq 1 for subsequent PLR analysis. We will discuss the differences in PLRs between these three sequences in Section \ref{dis}.

\subsection{Multi-Band Data}
$\delta$ Sct stars in Seq 1 were cross-matched with Gaia DR3 to obtain the parallax and photometry in the $G$, $BP$, and $RP$ bands, and we obtained parameters for 4,359 $\delta$ Sct stars \citep{2023A&A...674A...1G}. We further cross-matched this sample with the American Association of Variable Star Observers (AAVSO) Photometric All-Sky Survey (APASS) \citep{2014CoSka..43..518H}, the Two Micron All Sky Surve \citep[2MASS][]{2006AJ....131.1163S}, and the Wide-field Infrared Survey Explorer (WISE) \citep{2010AJ....140.1868W} catalogs for photometric data in the $V$, $J$, $H$, $K_S$, $W_1$, and $W_2$ bands. Due to insufficient sample in the $V$ band, we excluded it from further analysis. For the WISE bands, we retrieved data from both the ALLWISE and NEOWISE \cite{2011ApJ...731...53M} catalogs. Using optical-band periods, we phase-folded the WISE light curves and applied a fourth-order Fourier series fit to derive mean magnitudes in the $W_1$ and $W_2$ bands (details provided in \cite{2018ApJS..237...28C}). In the case of 2MASS, the reported magnitudes represent the mean of six consecutive 1.3-second exposures. For variable stars, this effectively corresponds to a single observational epoch, making it difficult to accurately derive phase-averaged magnitudes. Given that our key results are not critically dependent on the 2MASS system, we retain these values while acknowledging their limitations. Additionally, for PLRs, the statistical significance remains robust as long as the sample size is sufficiently large, even if some photometric data are sparsely sampled. We calculated extinctions using Gaia's \texttt{AG\_{GSPPHOT}} and determined the extinction values for each band according to coefficients published by \cite{2019ApJ...877..116W, 2023ApJ...956...26L}. Additionally, we derive optical and NIR Wesenheit indices \citep{1982ApJ...253..575M}. Wesenheit magnitudes are particularly convenient for calibrating the PLR since they are independent of reddening, it is commonly used as $W$ = $V$ - $R(I-V)$. We used Gaia's $G$, $BP$, and $RP$ and along with 2MASS' $J$, $H$, and $K_S$ passband and ALLWISE $W_1$, $W_2$ band to form the Wesenheit magnitudes as $G - 1.89 (BP - RP)$, $J - 0.686 (K_S - J)$, $K - 1.464 (H - K)$, $W_2 - 2.032 (W_1 - W_2)$ and $W_1 - 0.094 (BP - RP)$. The adopted coefficients are from the extinction coefficients.

\section{Method}\label{method}
We first corrected the Gaia parallax using the \texttt{zero-point} Python package \citep{lindegren2021gaiaads} and applied a cutoff by Corrected\_parallax $>$ 0. We also excluded $\delta$ Sct stars with renormalized unit weight error (RUWE) parameters exceeding 1.4 \citep{2018aa...616A...2L}.
To obtain the PLR, we determined both the potential zero-point deviation \(zp_{\varpi}\) of the Gaia parallax and the PLR coefficients in `parallax space' using the method proposed by \cite{2018ApJ...861..126R,2021ApJ...908L...6R}. The Gaia parallax is still biased after correction, and its value varies with versus magnitude, color, and spatial position. Ignoring this bias will result in a deviation in the PLR zero point. We applied the weighted nonlinear least squares method to fit the Equ. \ref{para}. The weights were estimated based on Equ. \ref{weight}, where $e_{\mathrm{plx}}$ is the parallax error, $\varpi$ is the parallax, and $\sigma_{m}$ is the magnitude error. 0.16 mag is $1\sigma$ intrinsic dispersion we assumed for the $\delta$ Sct PLR.

\begin{equation}
\varpi=1/10^{(0.2(m + a_1(\log P_{F}-\log P_{0}) +a_2)-2)} +z p_{\varpi}
    \label{para}
\end{equation}

\begin{equation}
    e_{\text {para }}^{2}= e_{\mathrm{plx}}^{2}+\varpi^{2} \times \frac{0.16^{2}+\sigma_{m}^{2}}{4 \times 1.086^{2}}\label{weight}
\end{equation}

In this method, when we used different parallax thresholds ($e_{\text {plx }}/\varpi$), it leads to different number of $\delta$ Sct stars and different final determined parallax \(zp_{\varpi}\) and PLR parameters. Therefore we need to determine the most appropriate parallax threshold by testing.

\begin{figure*}
	\begin{center}
\includegraphics[width=1.0\linewidth]{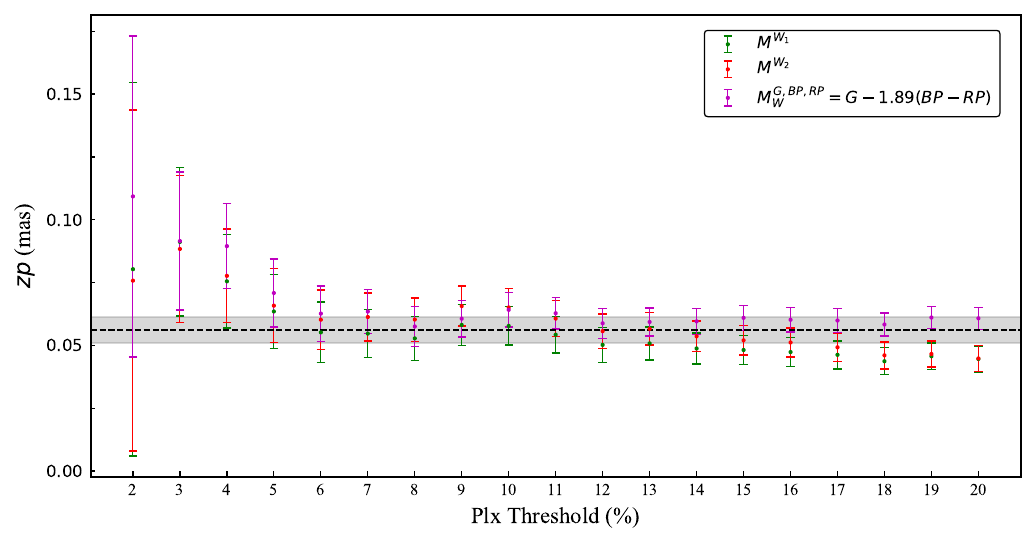}
		\caption{The parallax \(zp_{\varpi}\) values for three bands under different parallax uncertainty threshold $e_{\text {plx }}/\varpi$.}
		\label{zp}
	\end{center}
\end{figure*}

Taking into account the effects of extinction and the intrinsic dispersion of the PLR, we prioritized the use of the infrared \(W_1\), \(W_2\), and \(G - 1.89 (BP - RP)\) bands to determine the parallax zero point \(zp_{\varpi}\). We analyzed and determined the optimal \(zp_{\varpi}\) value by exploring various parallax uncertainty thresholds ranging from 2\% to 20\%. Fig. \ref{zp} presents the calculated \(zp_{\varpi}\) values for different parallax uncertainty thresholds in the three bands. The \(zp_{\varpi}\) values derived from the three bands exhibit remarkable consistency. Overall, \(zp_{\varpi}\) remains relatively stable across different parallax uncertainty thresholds, indicating that sample selection exerts minimal influence on the derived zero point. At the parallax uncertainty threshold of 4\%, the sample size was significantly reduced, resulting in larger uncertainties in the derived \(zp_{\varpi}\) value. Consequently, we excluded this threshold from our final calculations. Instead, the final \(zp_{\varpi}\) value was determined by averaging the results obtained across the three bands for parallax uncertainties ranging from 5\% to 20\%. The associated uncertainty was calculated as the average of the standard deviations for the three bands (shaded region in Fig. \ref{zp}). Thus, the final \(zp_{\varpi}\) value was determined to be \(0.0562 \pm 0.0051 \, \text{mas}\).

\begin{figure*}
	\begin{center}
\includegraphics[width=1.0\linewidth]{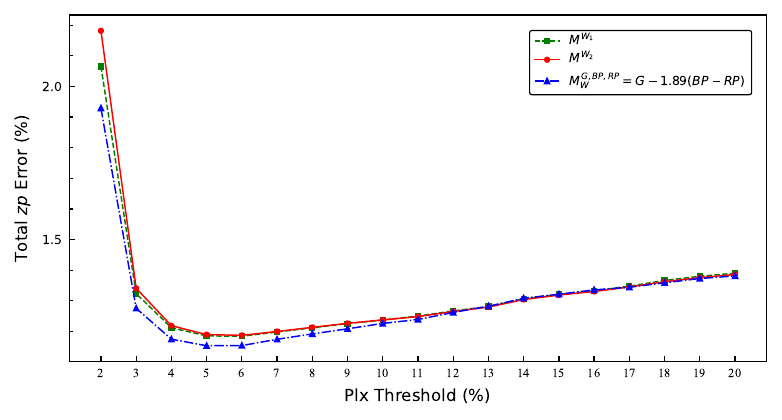}
		\caption{Parallax space total \(zp_{\varpi}\) error under different parallax uncertainty threshold $e_{\text {plx }}/\varpi$.}
		\label{th}
	\end{center}
\end{figure*}

After determining the parallax zero point \(zp_{\varpi}\), we need to determine a suitable parallax uncertainty threshold  ($e_{\text {plx }}/\varpi$) to make the PLR more accurate. Our idea is to minimize the total zero point error of PLR $tot_{err}$, which is calculated as Equ. \ref{para_error}. It includes fitting error and propagation error. Here, $\sigma_{a_2}$ is the error of PLR intercept. As shown in Fig. \ref{th}, the total zero point error first decreases as the parallax uncertainty threshold decreases. And when the parallax uncertainty threshold is less than 4\%, the increase in fitting error leads to the increase in total zero point due to the rapid decrease in the number of samples. We found that the total zero point error is minimized when the parallax uncertainty threshold is between 4\% and 7\%. We finally settled on 5\% as the parallax uncertainty threshold and used it for subsequent PLR calculations. The absolute magnitude of the star (\(M_\lambda\)) is derived from Equ. \ref{eabsmag}, which involves its apparent magnitude (\(m_\lambda\)), extinction (\(A_\lambda\)), and corrected parallax.

\begin{equation}
tot_{err} =\sqrt[2]{(\sigma_{a_2}/2/1.086)^{2}+ (5.1/mean(\overline{\omega})/1000)^{2}} 
    \label{para_error}
\end{equation}

\begin{equation}
M_{\lambda}=m_{\lambda}-5 \log \left(\varpi-z p_{\varpi}\right)+10-A_{\lambda}
\label{eabsmag}
\end{equation}

\section{Result}\label{result}
\begin{figure*}
	\begin{center}
\includegraphics[width=1.0\linewidth]{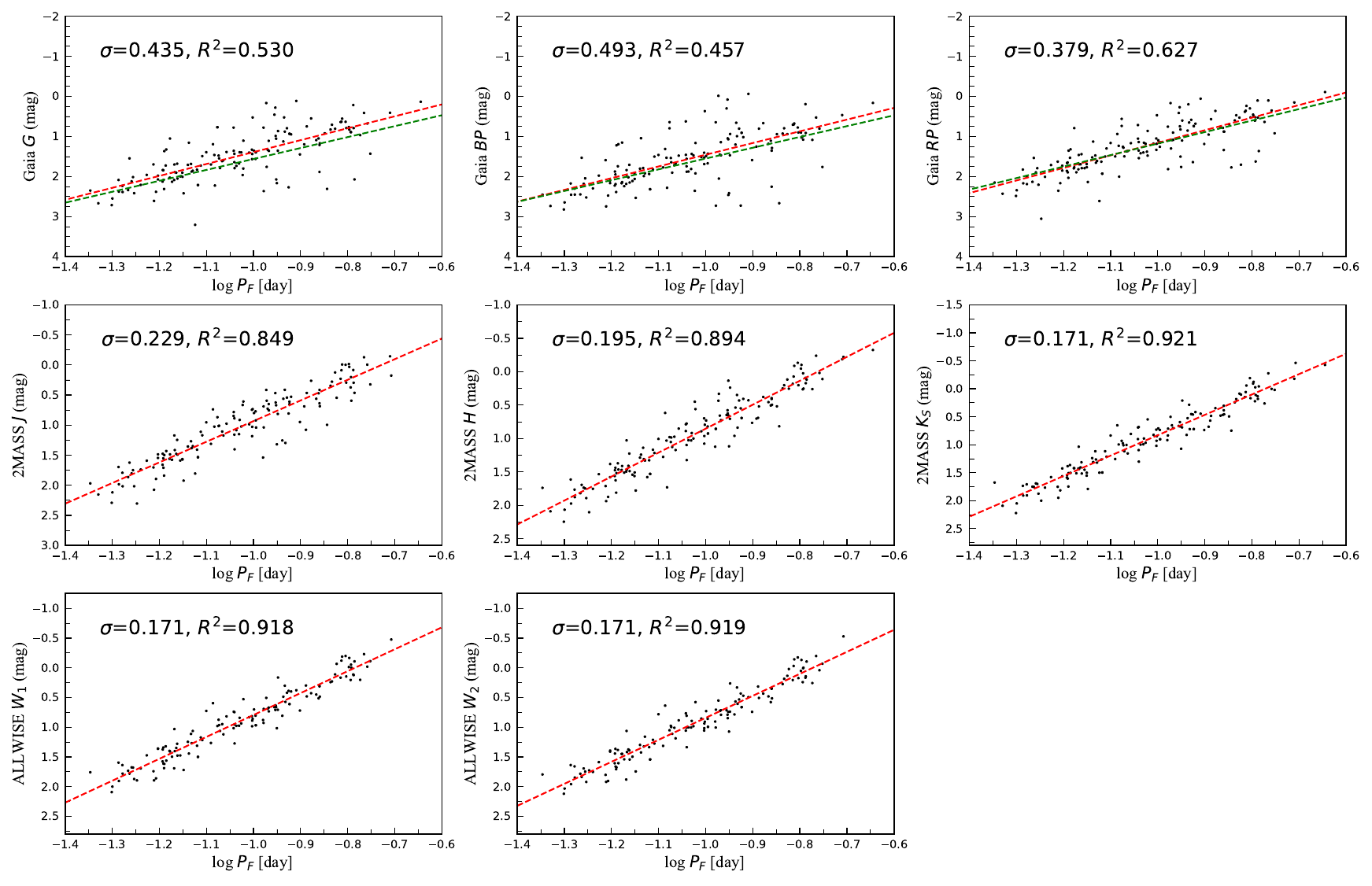}
		\caption{Eight-band PLRs for double-mode $\delta$ Sct stars by using F mode. The red dashed lines represent the fitted PLRs, while the PLRs derived from the LMC are shown as green dashed lines.}
		\label{plf}
	\end{center}
\end{figure*}

\begin{figure*}
	\begin{center}
\includegraphics[width=1.0\linewidth]{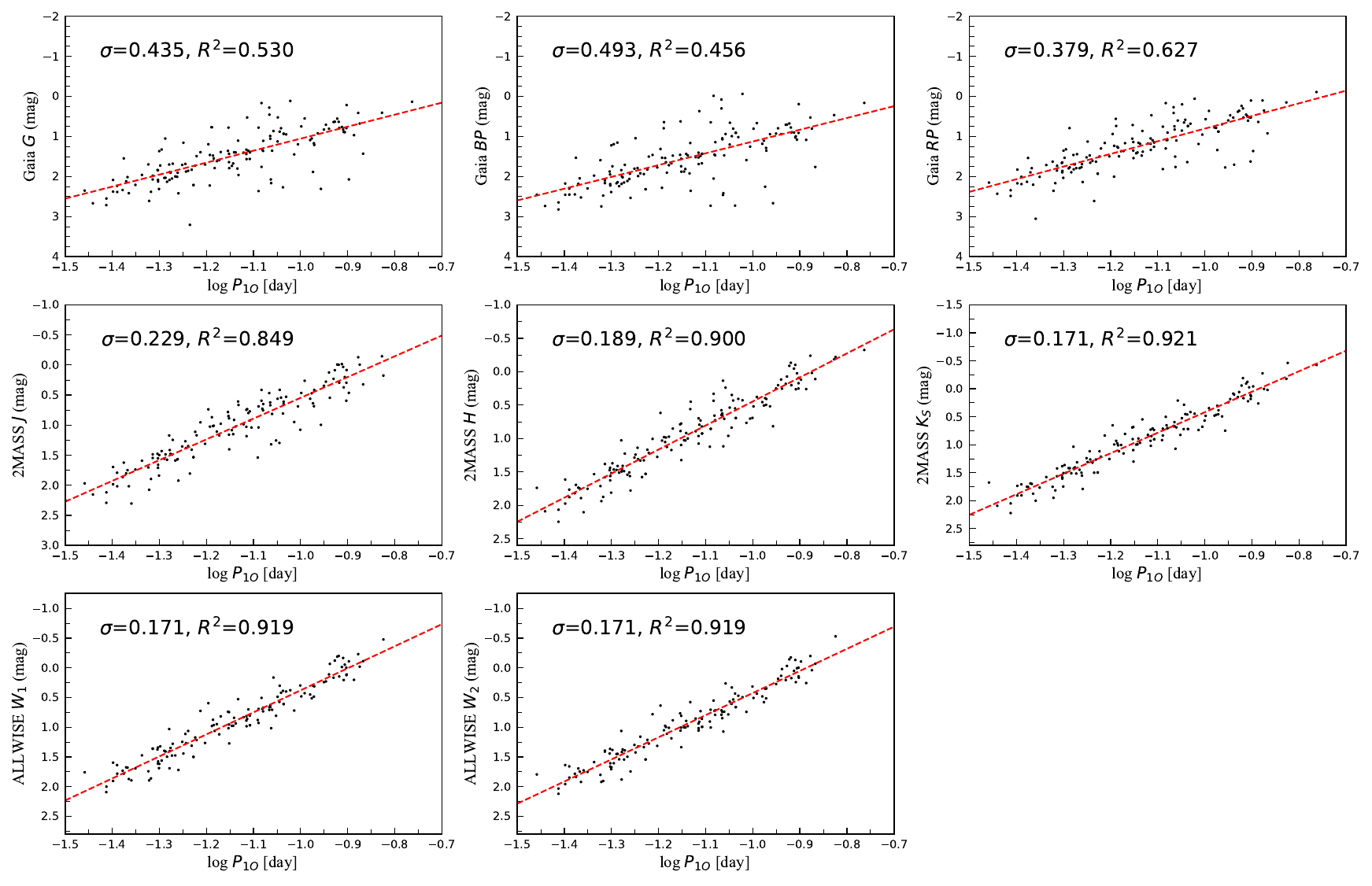}
		\caption{Eight-band PLRs for double-mode $\delta$ Sct stars using 1O mode.}
		\label{pl1O}
	\end{center}
\end{figure*}

\begin{figure*}
	\begin{center}
\includegraphics[width=1.00\linewidth]{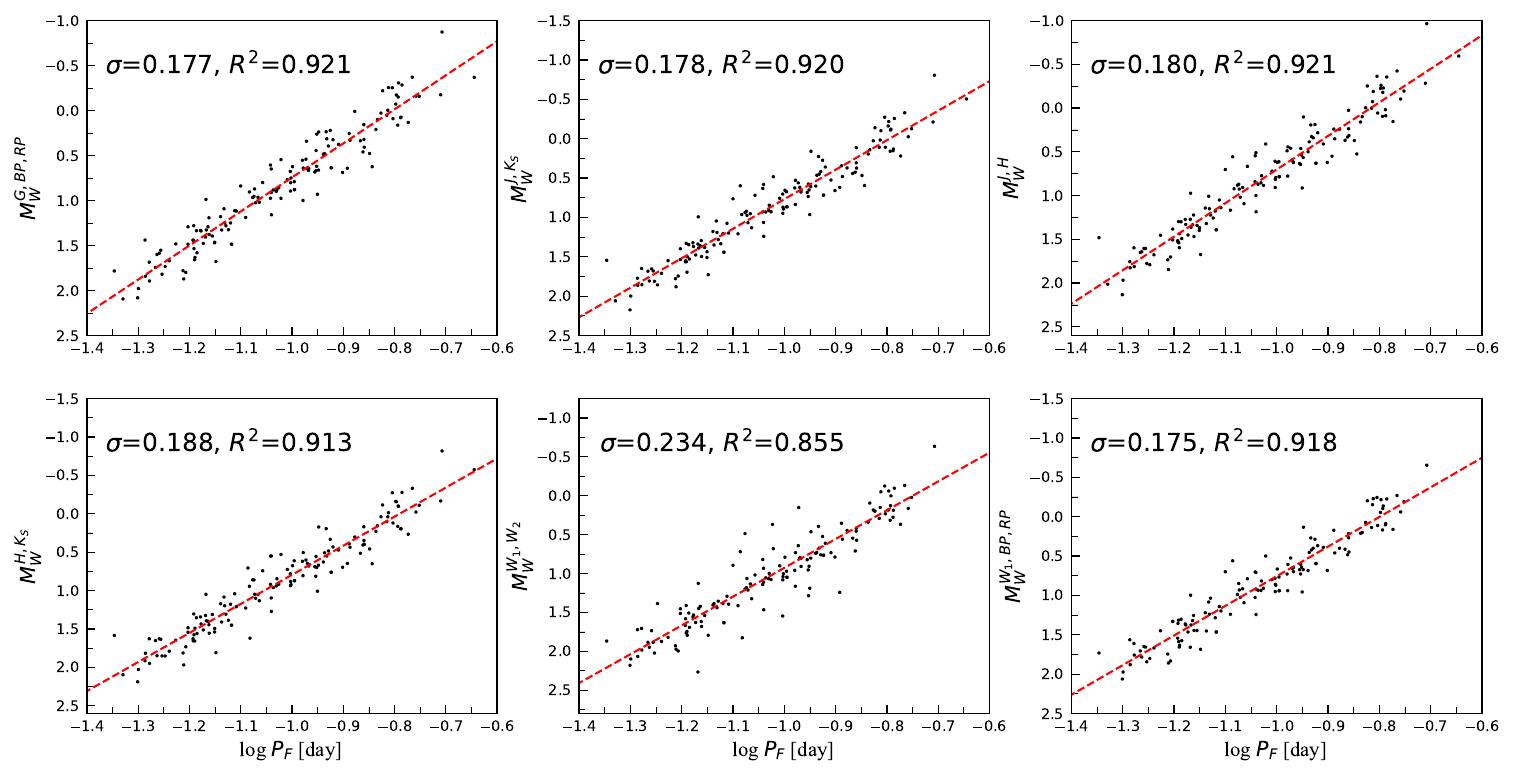}
		\caption{Six-band PW relations for double-mode $\delta$ Sct stars by using F mode.}
		\label{pwf}
	\end{center}
\end{figure*}

\begin{figure*}
	\begin{center}
\includegraphics[width=1.00\linewidth]{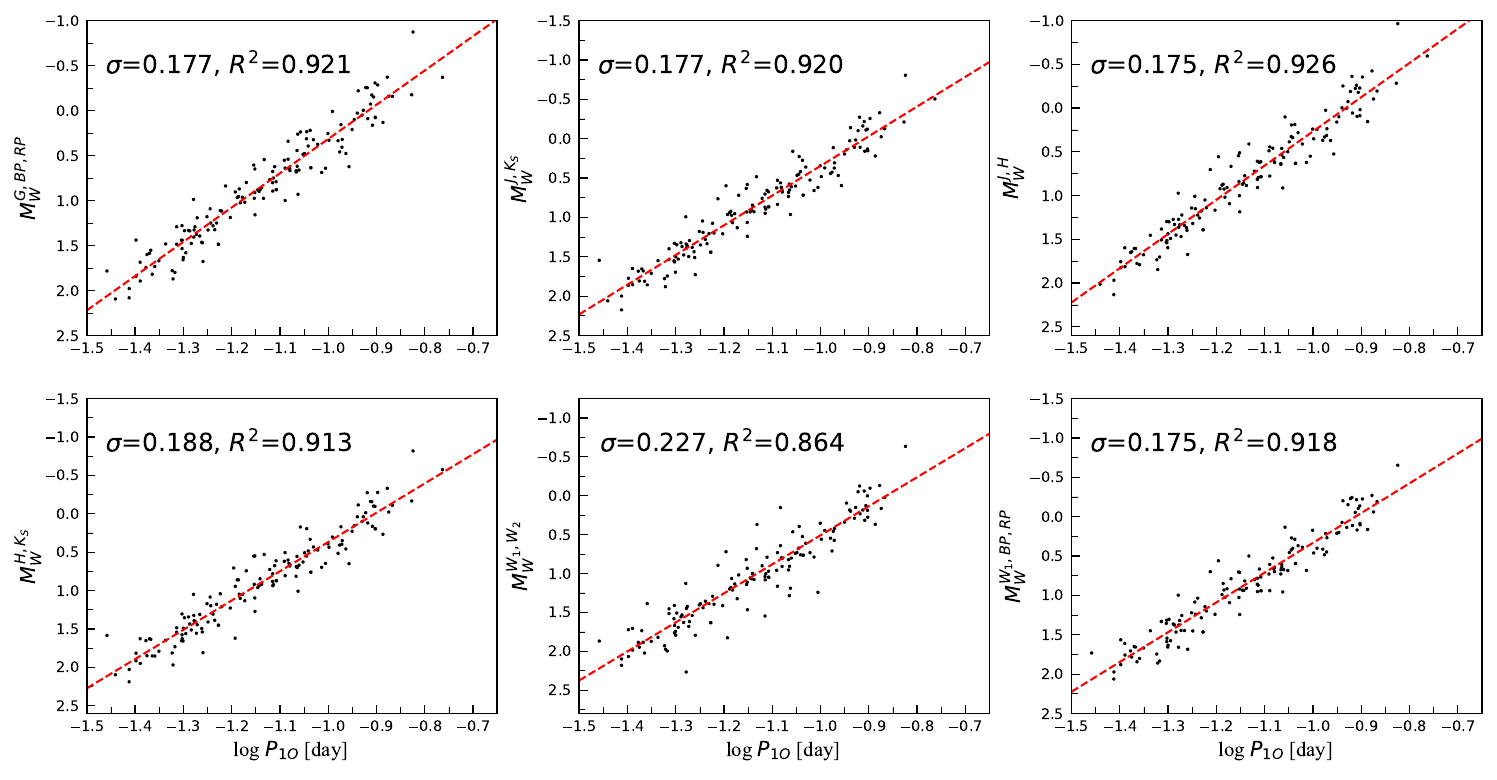}
		\caption{Six-band PW relations for double-mode $\delta$ Sct stars by using 1O mode.}
		\label{pw1O}
	\end{center}
\end{figure*}

We fitted PLRs in eight bands from optical to infrared using Equ. \ref{ePl},  where $\log{P}_{0}$ is the average period. In each fit, we first determined the PLR and eliminated data points with more than 3 $\sigma$, and the process was repeated three times. The results are shown in Fig. \ref{plf}, where the PLR has a small dispersion in the infrared band. Additionally, we analyzed the 1O modes of each $\delta$ Sct, as shown in Fig. \ref{pl1O}. We also fitted six PW relations for the F and 1O modes, as shown in Fig. \ref{pwf} and \ref{pw1O}.

\begin{equation}
M_{\lambda}=a_1 (\log{P}-\log{P}_{0})+ a_2 \label{ePl}
\end{equation}

We present the results of the PL and PW relations for double-mode $\delta$ Sct stars in Table \ref{plrparameter}, listing the eight bands ($G$, $BP$, $RP$, $J$, $H$, $K_S$, $W_1$, $W_2$) and six Wesenheit bands ($W_{G, BP, RP}$, $W_{J, K_S}$, $W_{H, K}$, $W_{W_1, W_2}$ and $W_{W_1, BP, RP}$). The number of double-mode $\delta$ Sct stars used to determine the PLR is about 140. Table \ref{filternumber} summarizes the selection steps for the PLR fitting sample and provides the number of $\delta$ Sct stars remaining at each step. The stringent parallax uncertainty criterion (5\%) resulted in the exclusion of over 90\% of the initial sample, significantly reducing the final dataset used for PLR fitting. Based on the analysis in Fig. \ref{th}, this sample is the most suitable for establishing the PLR. The dispersion $\sigma$ and total zero point error of the PLR decreases from the optical to the infrared band. In the $W1$ band, the dispersion is about 0.171 mag and the total zero point error is 1\%. This implies that the distance error obtained based on $\delta$ Sct star is about 8.7\%, whereas when the number of $\delta$ Sct stars is infinite, the distance error can be close to 1\%. Also in this band, $R^2$ is about 0.9, meaning that the PLR is well in line with the linear relationship. The coefficients of the 1O-mode PLRs differ very little from those of the F-mode after considering different average periods. For double-mode $\delta$ Sct, the period ratio of F mode and 1O mode is $\sim$ 0.774 \citep{2024ApJS..273....7J}, so the PLR between the 1O and F modes differs only by a constant related to the period ratio which has been taken into account in the average period item.

The LMC, as a nearby satellite galaxy of the Milky Way, serves as a fundamental anchor in the cosmic distance scale calibration. To validate the PLRs derived for Milky Way $\delta$ Sct stars, we extracted a subsample of LMC members from the Seq 1 dataset. Adopting a distance modulus of 18.477 mag for the LMC \citep{2019Natur.567..200P} and a mean reddening $E(V-I)$ = 0.10 $\pm$ 0.04 mag \citep{2021ApJS..252...23S}, we constructed PLRs in the $G$, $BP$, $RP$, as shown in Figures~\ref{plf} (green lines). The corresponding best-fit parameters are provided in Equation~\ref{dm}. Notably, the fitted LMC PLRs show good consistency with those derived from Galactic samples (Table~\ref{plrparameter}). We also emphasize several factors affecting the precision of the LMC-based relations: the typical apparent magnitude of these $\delta$ Sct stars is $G \sim 20$ mag, with photometric uncertainties of $\sim$0.01 mag in $G$ and up to $\sim$0.1 mag in the $BP$ and $RP$ bands. Moreover, the high stellar density in the LMC fields may introduce additional systematic uncertainties due to source crowding and blending, which could impact the accuracy of the derived PLRs.

Conversely, we can also use the $\delta$ Sct stars' PLR to derive the distance to the LMC. Since only optical bands are available, extinction dominates the uncertainty in the distance. Using the PLR in the $G$ band, we obtain an LMC distance of $18.74 \pm 0.02 \pm 0.09$ mag, where 0.02 mag is the statistical uncertainty and 0.09~mag corresponds to a 10\% uncertainty in extinction (mean extinction $A_G = 0.9$~mag for the Milky Way sample). This result differs from that of \citet{2019Natur.567..200P}  by $2.8\sigma$. In the future, with observations from infrared facilities such as James Webb Space Telescope (JWST), the LMC distance based on $\delta$~Sct stars will become more accurate.

\begin{equation}
\begin{aligned}
M_{G} &= (-2.728 \pm 0.080) (\log{P_F} + 0.942) + (1.562 \pm 0.010), \quad \sigma = 0.190 \, \text{mag}, \\
M_{BP} &= (-2.696 \pm 0.172) (\log{P_F} + 0.933) + (1.552 \pm 0.023), \quad \sigma = 0.385 \, \text{mag}, \\
M_{RP} &= (-2.864 \pm 0.178) (\log{P_F} + 0.934) + (1.178 \pm 0.024), \quad \sigma = 0.392 \, \text{mag} \\
\end{aligned}
\label{dm}
\end{equation}

\begin{deluxetable*}{ccccccccccc}
    \centering
\tablecaption{Double-mode $\delta$ Sct PLRs based on samples with 5\% parallax uncertainty criterion.}
    \label{plrparameter}
    \tablewidth{1pt}
    \tabletypesize{\small}
    \tablehead{
Mode&Passband& $a_1$&$\log{P}_{0}$&$a_2$&$tot_{err}$&$\sigma$&$R^2$&$N_{stars}$\\&&&(days)&&&}
\startdata
\multirow{14}{*}{F}&$G$&-2.974$\pm$0.236 & -1.032 & 1.486$\pm$0.037 & 0.019 & 0.435 & 0.530 & 143 \\&$BP$&
-2.921$\pm$0.270 & -1.033 & 1.553$\pm$0.042 & 0.021 & 0.493 & 0.457 & 141 \\&$RP$&
-3.135$\pm$0.202 & -1.031 & 1.254$\pm$0.032 & 0.017 & 0.379 & 0.627 & 146 \\&$J$&
-3.434$\pm$0.126 & -1.030 & 1.035$\pm$0.020 & 0.013 & 0.229 & 0.849 & 135 \\&$H$&
-3.584$\pm$0.102 & -1.025 & 0.943$\pm$0.016 & 0.012 & 0.195 & 0.894 & 147 \\&$K_S$&
-3.642$\pm$0.090 & -1.026 & 0.924$\pm$0.014 & 0.011 & 0.171 & 0.921 & 143 \\&$W_1$&
-3.690$\pm$0.096 & -1.029 & 0.903$\pm$0.015 & 0.011 & 0.171 & 0.918 & 133 \\&$W_2$&
-3.703$\pm$0.096 & -1.028 & 0.948$\pm$0.015 & 0.011 & 0.171 & 0.919 & 132 \\&$G, BP, RP$&
-3.783$\pm$0.093 & -1.024 & 0.835$\pm$0.015 & 0.010 & 0.177 & 0.921 & 145 \\&$J, K_S$&
-3.746$\pm$0.093 & -1.026 & 0.865$\pm$0.015 & 0.010 & 0.178 & 0.920 & 143 \\&$J, H$&
-3.832$\pm$0.095 & -1.026 & 0.801$\pm$0.015 & 0.010 & 0.180 & 0.921 & 143 \\&$H, K_S$&
-3.786$\pm$0.099 & -1.025 & 0.891$\pm$0.016 & 0.010 & 0.188 & 0.913 & 143 \\&$W_1, W_2$&
-3.702$\pm$0.131 & -1.028 & 1.027$\pm$0.020 & 0.012 & 0.234 & 0.855 & 137 \\&$W_1, BP, RP$&
-3.760$\pm$0.098 & -1.029 & 0.868$\pm$0.015 & 0.010 & 0.175 & 0.918 & 133 \\
\hline
\multirow{14}{*}{1O}&$G$&-2.992$\pm$0.237 & -1.144 & 1.486$\pm$0.037 & 0.018 & 0.435 & 0.530 & 143 \\&$BP$&
-2.940$\pm$0.272 & -1.145 & 1.553$\pm$0.042 & 0.020 & 0.493 & 0.456 & 141 \\&$RP$&
-3.154$\pm$0.203 & -1.142 & 1.254$\pm$0.032 & 0.016 & 0.379 & 0.627 & 146 \\&$J$&
-3.454$\pm$0.126 & -1.142 & 1.035$\pm$0.020 & 0.011 & 0.229 & 0.849 & 135 \\&$H$&
-3.599$\pm$0.100 & -1.137 & 0.938$\pm$0.016 & 0.010 & 0.189 & 0.900 & 146 \\&$K_S$&
-3.666$\pm$0.090 & -1.138 & 0.924$\pm$0.014 & 0.009 & 0.171 & 0.921 & 143 \\&$W_1$&
-3.710$\pm$0.097 & -1.141 & 0.903$\pm$0.015 & 0.009 & 0.171 & 0.919 & 133 \\&$W_2$&
-3.724$\pm$0.097 & -1.140 & 0.948$\pm$0.015 & 0.009 & 0.171 & 0.919 & 132 \\&$G, BP, RP$&
-3.807$\pm$0.093 & -1.136 & 0.835$\pm$0.015 & 0.010 & 0.177 & 0.921 & 145 \\&$J, K_S$&
-3.770$\pm$0.094 & -1.138 & 0.865$\pm$0.015 & 0.010 & 0.177 & 0.920 & 143 \\&$J, H$&
-3.908$\pm$0.093 & -1.135 & 0.796$\pm$0.015 & 0.010 & 0.175 & 0.926 & 142 \\&$H, K_S$&
-3.810$\pm$0.099 & -1.137 & 0.891$\pm$0.016 & 0.010 & 0.188 & 0.913 & 143 \\&$W_1, W_2$&
-3.733$\pm$0.128 & -1.139 & 1.027$\pm$0.020 & 0.011 & 0.227 & 0.864 & 136 \\&$W_1, BP, RP$&
-3.781$\pm$0.099 & -1.141 & 0.868$\pm$0.015 & 0.010 & 0.175 & 0.918 & 133 \\
\enddata
\tablecomments{Mode: double-mode $\delta$ Sct star pulsation mode; Passband: photometric data band used; $a_1$: PLR slope; $\log{P}_{0}$: the mean value of the F mode period or the 1O mode period; $a_2$: PLR intercept; $tot_{err}$: total zero point error, calculated from Equation (\ref{para_error}); $R^2$: fit goodness-of-fit values; $\sigma$: PLR dispersion; $N_{stars}$: number of stars for fitting.}
\end{deluxetable*}

\begin{deluxetable*}{lcccccc}
\centering
\tablecaption{Sample Selection Process for $\delta$ Sct Star PLR determination.}
    \label{filternumber}
\tablewidth{1pt}
\tabletypesize{\small}
\tablehead{
    \colhead{Selection Step} & 
    \colhead{Retained Stars} &
    \colhead{MW Stars} & 
    \colhead{LMC Stars} & 
    \colhead{SMC Stars} }
\startdata
Initial sample & 6426 & 5907 & 477 & 42  \\
Sequence selection (Seq1) & 4539 & 4148 & 382 & 9 \\
Passband data availability & 4499 & 4135 & 357 & 7  \\
Corrected\_parallax $>$ 0, RUWE $<$ 1.4, $m_{\lambda}$ $>$ 0 & 2847 & 2736 & 109 & 2  \\
5\% parallax uncertainty & 151 & 150 & 1 & 0  \\
3$\sigma$ outlier rejection & 143 & 142 & 1 & 0  
\enddata
\tablecomments{MW: Milky Way, LMC: Large Magellanic Cloud, SMC: Small Magellanic Cloud.}
\end{deluxetable*}

\section{discussion}
\label{dis}
\subsection{Sequence affect}
This section discusses the differences in PLR between the three sequences on the Petersen diagram (see Section \ref{petersen detail}). We begin with an initial analysis of all samples to obtain $\sigma$ and $R^2$ (as shown in Table \ref{allseq}). We find that the primary sequence Seq 1 has smaller $\sigma$, which prompted our decision to use Seq 1 alone to obtain the PLRs. Seq 1 as the primary sequence is the typical double-mode $\delta$ Sct, the relative description can also be found in \cite{2020AcA....70..101S, 2021AcA....71..189S, 2022AcA....72..245S, 2023AcA....73..105S}

\begin{deluxetable*}{cccccccccccccccc}
    \centering
\tablecaption{PLR dispersion differences between all samples and Seq 1 samples.}
    \label{allseq}
    \tablewidth{1pt}
    \tabletypesize{\small}
\tablehead{\colhead{\multirow{2}{*}{Band}}&\multicolumn{2}{c}{$\sigma$}&&\multicolumn{2}{c}{$R^2$}\\
\cline{2-3}\cline{5-6}
\colhead{} &\colhead{all} & \colhead{Seq 1} & &\colhead{all} & \colhead{Seq 1}}
\startdata
$G$&0.563 & 0.435 &&0.392&0.530 \\
$BP$&0.638 & 0.493&&0.317&0.457\\
$RP$&0.462 & 0.379 && 0.507&0.627\\
$J$&0.288& 0.229&& 0.797&0.849\\
$H$&0.212 & 0.195&& 0.889&0.894\\
$K_S$&0.220 & 0.171&& 0.888&0.921\\
$W_1$&0.222 & 0.171&&0.888&0.918\\
$W_2$&0.217 & 0.171&&0.891&0.919\\
\enddata
\end{deluxetable*}

\begin{figure*}
	\begin{center}
\includegraphics[width=1.0\linewidth]{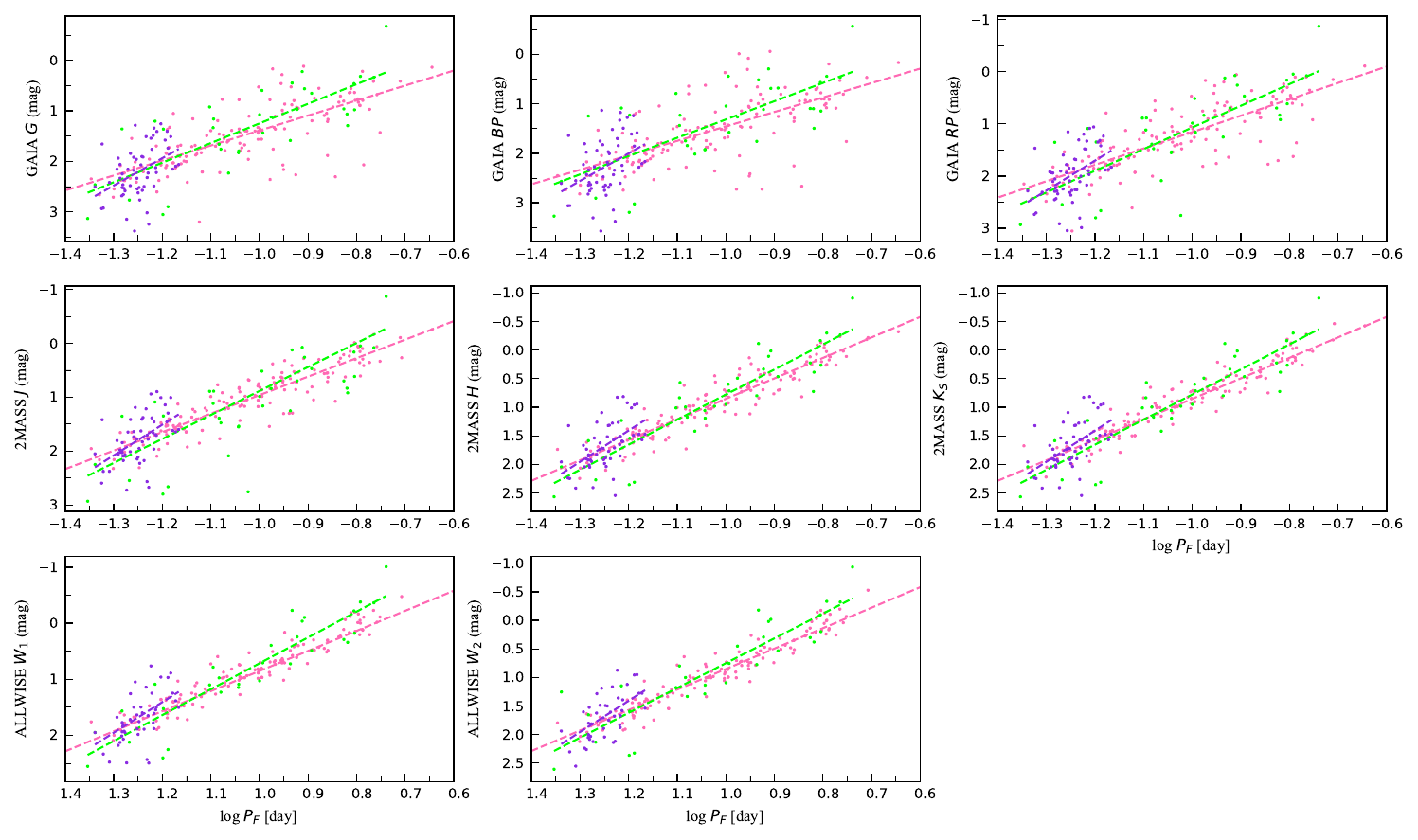}
		\caption{Eight-band PLRs for Seq 1, 2 and 3. The pink color represents Seq 1, the purple color represents Seq 2, and the green color represents Seq 3.}
		\label{plrseq123}
	\end{center}
\end{figure*}

In Seq 2, because the objects are much farther away, using the same parallax uncertainty criterion as in Seq 1 left only a few objects. We adjusted the parallax criterion several times and found that 20\% uncertainty produced the best fit, which had the similar PLRs as Seq 1 (Fig. \ref{plrseq123}).

\begin{figure*}
	\begin{center}
\includegraphics[width=1.0\linewidth]{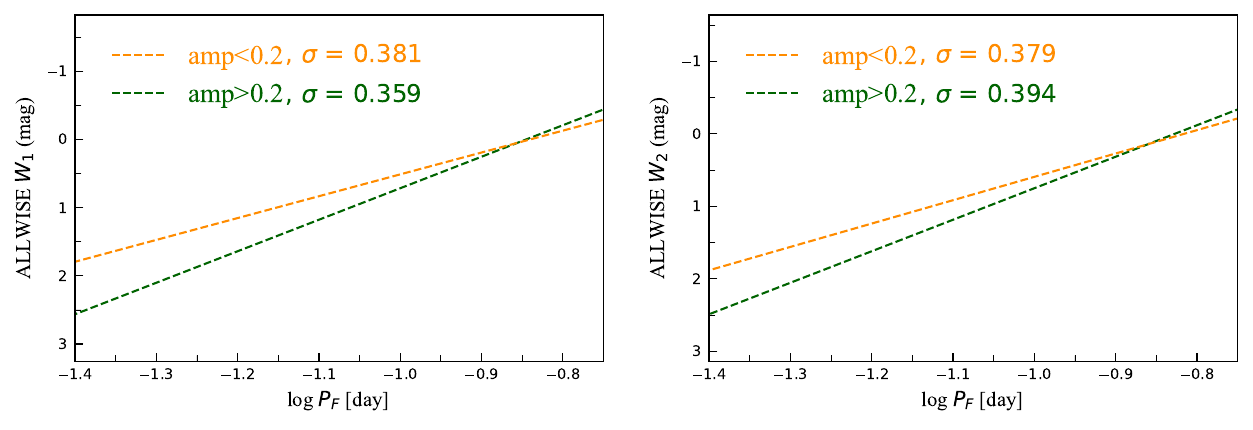}
		\caption{Comparison of PLRs for samples with different amplitudes in Seq 3, with the darkorange line representing amplitude less than 0.2 mag and the darkgreen line representing amplitude greater than 0.2 mag.}
		\label{amp2}
	\end{center}
\end{figure*}

For Seq 3, we found the PLR slope to be significantly smaller than the other sequences. The difference in slope could not be mitigated by adjusting the parallax uncertainty criterion. The difference is mainly due to the purity of double-mode $\delta$ Sct, which is more dispersed on the Petersen diagram for Seq 3 compared to Seq 1 and 2. Objects on Seq 3 may not be a typical double-mode $\delta$ Sct when the two periods correspond to smaller amplitudes, i.e., the periods are not the radial F and 1O modes. For this reason, we compared the PLRs for samples with amplitudes greater than 0.2 mag and less than 0.2 mag (shown in Fig. \ref{amp2}) and found that the slope of the PLRs with amplitudes greater than 0.2 mag is larger and more consistent with the other sequences (as shown in Fig. \ref{plrseq123}). Thus for Seq 3, the accuracy of the PLR can be improved by using amplitudes greater than 0.2 mag as a criterion to ensure the sample purity.

\begin{figure*}
	\begin{center}
\includegraphics[width=1.0\linewidth]{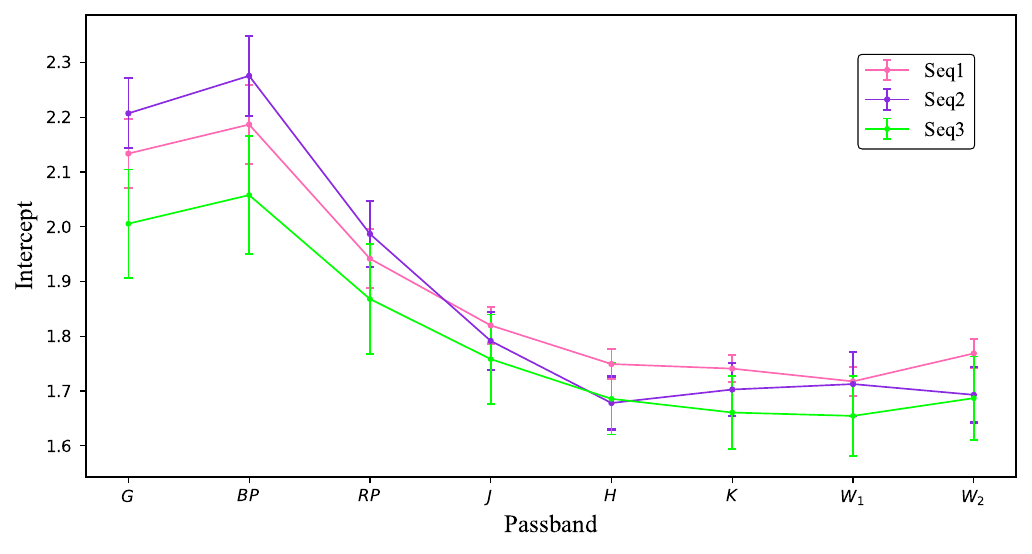}
		\caption{Distribution of PLR's zero points at $\log P_0 = -1.25$ in eight bands for Seq 1, 2 and 3. The pink color represents Seq 1, the purple color represents Seq 2, and the green color represents Seq 3.}
		\label{intercept123}
	\end{center}
\end{figure*}

We chose $\log P_0 = -1.25$ to visually compare the difference in the PLR zero points of the three sequences, and we fixed the slopes all to the slope of the PLR of Seq 1 and fitted the intercepts of the three sequences. In Fig. \ref{intercept123}, we compare the intercepts of the three sequences over different bands and find that they are consistent after accounting for errors.

The similarity in the PLRs of the three sequences suggests that the luminosity and evolutionary properties of these double-mode $\delta$ Sct are similar. Compared to Seq 1, the Seq 2 sample has larger parallax errors and a narrower period range, which tends to bias the PLRs. The Seq 3 sample requires more careful identification of F and 1O modes. Therefore, we recommend using the Seq 1 sample to determine the PLR.

\subsection{Comparison with the Literature}
\begin{figure*}
	\begin{center}
\includegraphics[width=1.0\linewidth]{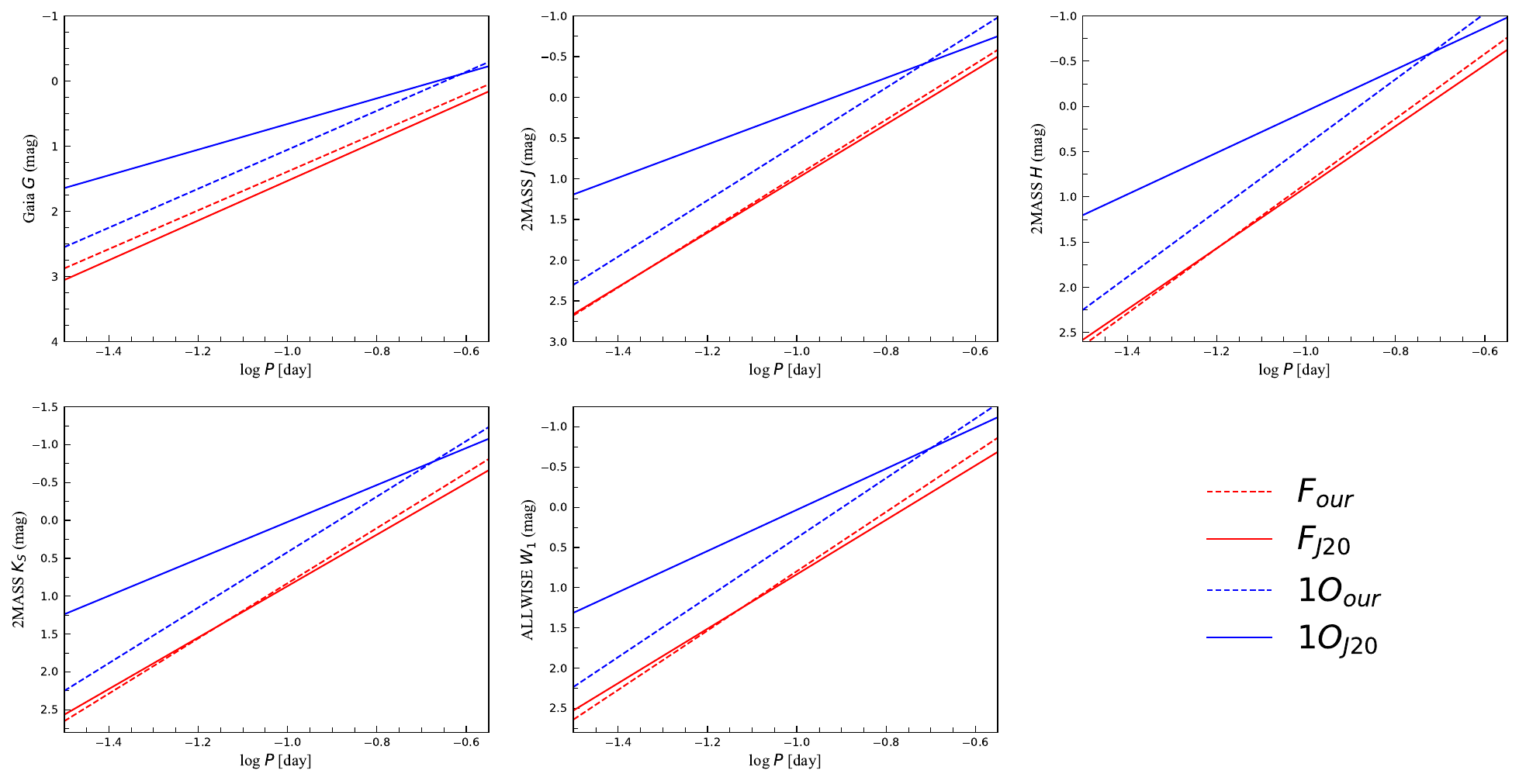}
		\caption{Comparison of the PLRs for $\delta$ Sct stars across five bands with J20. The blue dashed and red dotted lines represent the results of our fitted F and 1O modes, respectively, while the blue solid and red solid lines correspond to the F and 1O mode PLRs from J20.}
		\label{vs}
	\end{center}
\end{figure*}

We compared our obtained results with those obtained by \cite{J20} (hereafter J20), as shown in Fig. \ref{vs}. We find that our results are similar to theirs in the F mode, but our results are significantly different from theirs in the 1O mode. This is due to the differences in determining mode classification between the two works. Our and OGLE's samples were obtained by a strict selection on the Petersen diagram \citep{2024ApJS..273....7J, 2020AcA....70..101S}. As a result, our 1O and F PLRs nearly parallel. In J20's work, they used PL diagram for classification, and after identifying samples with F modes, they considered all samples above the line of fit of the F mode PLR as 1O modes. This classification approach leads to mode confusion, with more higher order pulsation modes being recognized as the 1O mode. This can lead to smaller PLR slopes and biases in the PLRs. 

The PLR we obtained in the infrared band demonstrates less dispersion than the results reported by J20, and only two bands show greater dispersion than theirs, as detailed in Table \ref{comparejasinghe}. However, it is noteworthy that the sample size we used is less than one-tenth of theirs. The PLRs derived from double-mode $\delta$ Sct stars are better than those from single-mode $\delta$ Sct stars. Unlike Cepheids and RR Lyrae, higher-order pulsation modes and non-radial pulsations are prevalent in $\delta$ Sct stars, and thus pulsation mode identification in $\delta$ Sct needs to be done very carefully.

\begin{deluxetable*}{cccccccccc}
\tablecaption{Six-band PLR dispersions compare with J20\label{comparejasinghe}.}

\tablehead{\colhead{\multirow{2}{*}{Mode}}& \colhead{\multirow{2}{*}{Band}}&&\multicolumn{2}{c}{$\sigma$}&&\multicolumn{2}{c}{$N_{stars}$}\\
\cline{4-5}\cline{7-8}
\colhead{} &\colhead{} &&\colhead{J20} & \colhead{This work} & &\colhead{J20} & \colhead{This work}}
\startdata
\multirow{6}{*}{F} &$G$  && 0.210 &0.435  &&1408&143\\
& $J$ && 0.188  & 0.229  &&1400&135\\
&$H$ & & 0.185  & 0.195& &1393 &147\\
&$K_S$ &&0.183  & 0.171 & & 1391 &143&\\
&$W_1$ & &0.190 & 0.171 & &1406 &133\\
&$J, K_S$& &0.184  & 0.178 & & 1396&143&
\\
\hline
\multirow{6}{*}{1O} & $G$ & &0.263  & 0.435 &  &1408&143&\\
&$J$& &0.223 & 0.229 &  &1400&135&\\
&$H$  &  &0.213&  0.189 & &1393&146&\\
&$K_S$  & & 0.209& 0.171 &  &1391&143 &\\
&$W_1$  & &0.218  & 0.171 & &1406 &133&\\
&$J, K_S$& &  0.209& 0.177 & & 226&143&
\enddata
\tablecomments{$\sigma$: PLR dispersion; $N_{stars}$: number of stars for fitting.}
\end{deluxetable*}

\subsection{Metalicity affect}
\begin{figure*}
	\begin{center}
\includegraphics[width=1.0\linewidth]{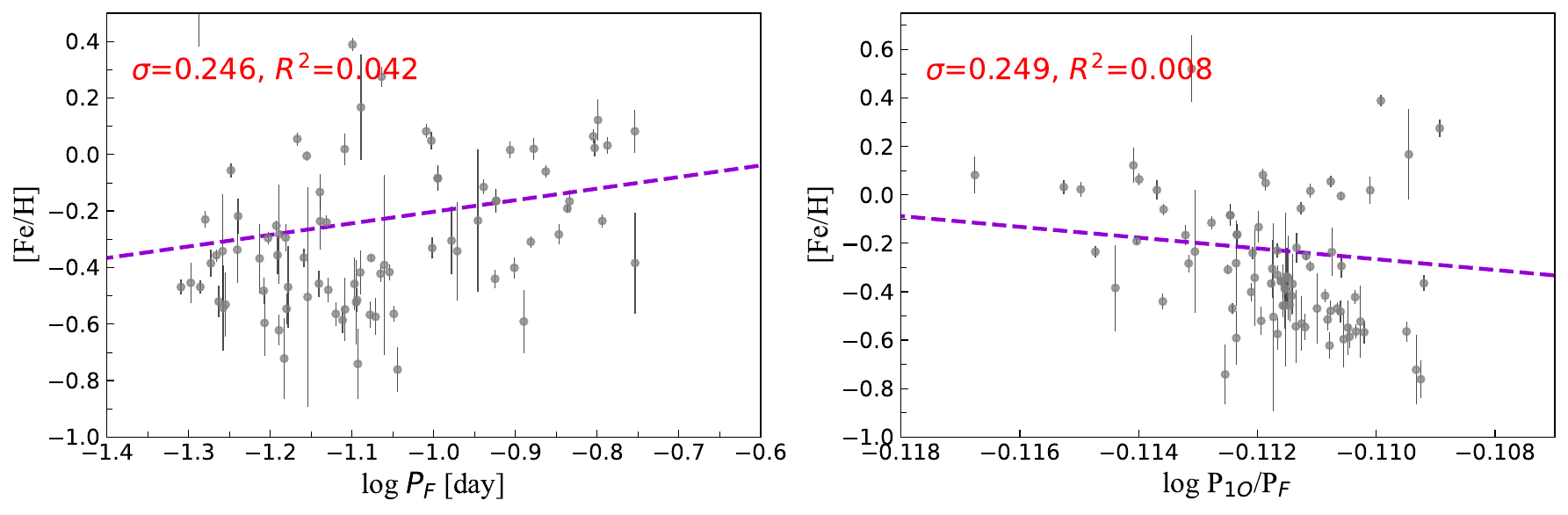}
		\caption{Period-metallicity diagram and period ratio-metallicity diagram of double-mode $\delta$ Sct stars.}
		\label{feh}
	\end{center}
\end{figure*}

In the work of \cite{2023NatAs...7.1081C}, they found a good linear relationship between metallicity and period as well as the metallicity to period ratio for double-mode RR Lyrae stars. They proposed that RRd stars are robust distance tracers whose distance measurements can be made by measuring only the period and can avoid being affected by metallicity. Here, we similarly explore whether double-mode $\delta$ Sct stars has similar properties. We obtained a sample of 88 double-mode $\delta$ Sct stars on Seq 1 with metallicity data by crossing LAMOST DR11 and determined the relation between period and metallicity. Their atmospheric parameters data are shown in Table \ref{lamostdr11}. To evaluate the correlation between metallicity and period or period ratio, we applied a weighted average fit. The results, shown in Fig. \ref{feh}, show that the coefficient of determination for both fits is very low ($R^2 < 0.1$). Additionally, the unweighted results also show a low correlation. These indicate that the period and metallicity of the double-mode $\delta$ Sct are independent. Unlike Cepheids and RR Lyrae, the metallicity effect of $\delta$ Sct PLR is similarly weak, i.e., both luminosity and period correlate weakly with metallicity. In another recent work, we investigated the metallicity effect of the PLR in the F mode of single period $\delta$ Sct and found that the effect was also not significant ($R^2 \sim 0.1$). We suspect that this is mainly due to the higher temperature of $\delta$ Sct, which has fewer spectral lines for its metal.

\begin{deluxetable*}{cccccccccccccccc}
    \centering
\tablecaption{Stellar metallicity and physical parameter data from LAMOST DR11 for Seq1's $\delta$ Sct stars.}
    \label{lamostdr11}
    \tablewidth{1pt}
    \tabletypesize{\small}
\tablehead{ R.A.(J2000) & Decl.(J2000)  & F & 1O & ...  & Teff    & $\log g$ & ...& [Fe/H]\\
(deg)&(deg)&(days)&(days)&...&($K$)&&...&}
\startdata
4.0690587   & 12.3890231 & 4.0690587   & 12.3890231    & ...  & 8728.5  & 4.538 & ...  &-0.469 \\
11.3423877  & 47.2456022 & 11.3423877  & 47.2456022    & ...  & 7178.51 & 4.017 & ...  &-0.156 \\
17.2227916  & 37.8402206 & 17.2227916  & 37.8402206    & ...  & 7129.25 & 4.013 & ...  &-0.548 \\
24.7461508  & 43.5562447 & 24.7461508  & 43.5562447    & ...  & 7243.9  & 4.067 & ...  &-0.309 \\
43.9162797  & 31.7551976 & 43.9162797  & 31.7551976    & ...  & 7223.43 & 4.257 & ...  &-0.591 \\
52.5173322  & 50.0381607 & 52.5173322  & 50.0381607    & ...  & 7212.91 & 4.025 & ...  &-0.166 \\
52.9199033  & 21.0217514 & 52.9199033  & 21.0217514    & ...  & 7370.04 & 4.12  & ...  &-0.722 \\
57.291309   & 5.6790994  & 57.291309   & 5.6790994     & ...  & 8659.99 & 4.368 & ...  &-0.23  \\
58.9839308  & 36.3880753 & 58.9839308  & 36.3880753    & ...  & 7207.0  & 4.317 & ...  &-0.458 \\
59.2722309  & 24.939244  & 59.2722309  & 24.939244     & ...  & 7318.48 & 3.961 & ...  &0.032  \\
60.6146648  & 55.3426712 & 60.6146648  & 55.3426712    & ...  & 7254.56 & 4.045 & ...  &-0.331 \\
67.222411   & 43.2530068 & 67.222411   & 43.2530068    & ...  & 7111.56 & 3.999 & ...  &-0.06  \\
67.5185315  & -6.1448143 & 67.5185315  & -6.1448143    & ...  & 7372.07 & 4.114 & ...  &-0.296 \\
74.1713386  & 12.0837865 & 74.1713386  & 12.0837865    & ...  & 7427.65 & 4.188 & ...  &-0.417 \\
79.9824919  & 33.0398638 & 79.9824919  & 33.0398638    & ...  & 8877.85 & 4.754 & ...  &-0.685 \\
86.3598859  & 12.2270466 & 86.3598859  & 12.2270466    & ...  & 7549.33 & 4.121 & ...  &-0.531 \\
95.8104118  & 13.3997492 & 95.8104118  & 13.3997492    & ...  & 7151.27 & 3.951 & ...  &-0.133 \\
98.2191896  & 16.1438321 & 98.2191896  & 16.1438321    & ...  & 7109.89 & 4.164 & ...  &-0.384 \\
98.7949174  & 56.507795  & 98.7949174  & 56.507795     & ...  & 7241.36 & 4.203 & ...  &-0.586 \\
99.0323963  & 20.4373335 & 99.0323963  & 20.4373335    & ...  & 7051.47 & 4.078 & ...  &-0.115 \\
99.6767199  & -1.0401371 & 99.6767199  & -1.0401371    & ...  & 7185.58 & 3.968 & ...  &-0.164 \\
103.5314181 & 5.0762254  & 103.5314181 & 5.0762254     & ...  & 7752.94 & 4.002 & ...  &-0.084 \\
104.4305676 & 23.6202137 & 104.4305676 & 23.6202137    & ...  & 7305.92 & 3.98  & ...  &-0.005 \\
... & ... & ... & ...    & ...  & ... & ... & ...&... \\
108.8115162 & 11.0350597 & 108.8115162 & 11.0350597    & ...  & 7105.89 & 4.244 & ...&-0.567 
\enddata
\tablecomments{R.A. and decl.: source position (J2000); F: fundamental period, 1O: first-overtone period, Teff: Effective temperature obtained by the LASP, $\log g$: Surface gravity obtained by the LASP, [Fe/H]: Metallicity obtained by the LASP.\\ LASP: The LAMOST Stellar Parameter Pipeline.\\(This table is available in its entirety in machine-readable form.)}
\end{deluxetable*}

\section{conclusions}
\label{conclusion}
In this study, we have conducted a comprehensive investigation of the PLRs for double-mode $\delta$ Sct stars, utilizing an all-sky sample derived from ZTF and OGLE, complemented by Gaia parallaxes and LAMOST metallicity data. Through parallax-space consistency checks, we derived a Gaia parallax zero point of $zp = 0.05620 \pm 0.0051 \, \text{mas}$, with an optimal parallax uncertainty threshold of 5\%, enabling a robust analysis of the PLRs.

Our results demonstrate that the PLRs in the infrared bands exhibit significantly lower dispersion compared to the optical bands, with the $W_1$ band showing the smallest total error ($tot_{\text{err}}$) of approximately 1\% and scatter of 0.171 mag. The PLRs of the 1O mode are nearly parallel to those of the F mode, with a small constant offset that correlates with the period ratio of the two modes. A comparison with previous studies reveals that our derived PLRs exhibit reduced dispersion, highlighting the critical importance of accurate mode classification in obtaining precise PLRs.

We also confirm the consistency of the PLRs across different sequences on the Petersen diagram, particularly after rigorous sample filtering. This consistency suggests that the luminosity and evolutionary properties of these $\delta$ Sct stars are similar.  We investigated the potential impact of metallicity on the PLRs of double-mode $\delta$ Sct stars. Metallicity is found to be weakly correlated with period and period ratio, unlike double-mode Cepheids and RR Lyrae stars.

Our findings underscore the utility of double-mode $\delta$ Sct stars for distance measurements within the Milky Way. The calibration of their PLRs, particularly in the infrared bands, offers significant potential for improving the accuracy of local distance scales and deepening our understanding of stellar evolution. While Cepheid variables are primary distance indicators, their PLR zero-point and metallicity effects remain uncertain. Future observations with JWST, Legacy Survey of Space and Time (LSST), and China Space Station Telescope (CSST) will identify more $\delta$ Sct stars in the Local Group, allowing independent distance measurements, particularly for the LMC, to refine the Cepheid PLR. Although $\delta$ Sct stars are intrinsically fainter, their greater numbers and intermediate-age population make them valuable for testing stellar population effects in distance determinations.

\section{Acknowledgements}
We thank the anonymous referee for the helpful comments. This research was supported by the National Natural Science Foundation of China (NSFC) through grants 12173047, 12322306, 12373028, 12233009, 12133002. X. Chen and S. Wang acknowledge support from the Youth Innovation Promotion Association of the Chinese Academy of Sciences (CAS, No. 2022055 and 2023065). We also thanked the support from the National Key Research and development Program of China, grants 2022YFF0503404. We used observations obtained the the Samuel Oschin Telescope 48-inch and the 60-inch Telescope at the Palomar Observatory as part of the ZTF project. ZTF is supported by the National Science Foundation under Grants No. AST-1440341 and AST-2034437. The authors gratefully acknowledge the use of highly valuable publicly available archival data from OGLE (\url{https://ogle.astrouw.edu.pl}). The OGLE project at the Las Campanas Observatory has been funded by the European Research Council under the European Community’s Seventh Framework Programme (FP7/2007-2013), ERC grant agreement no. 246678 awarded to AU. We used data from European Space Agency mission Gaia (\url{https://www.cosmos.esa.int/gaia}), processed by the Gaia Data Processing and Analysis Consortium (DPAC, \url{ https://www. cosmos.esa.int/web/gaia/dpac/consortium}). Funding for the DPAC has been provided by national institutions, in particular the institutions participating in the Gaia Multilateral Agreement. This work has made use of LAMOST data. Guoshoujing Telescope (LAMOST; \url{http://www.lamost.org/public/}) is a National Major Scientific Project built by the Chinese CAS. Funding for the project has been provided by the National Development and Reform Commission. LAMOST is operated and managed by the National Astronomical Observatories, CAS.

\bibliography{PLDMDSCT}
\bibliographystyle{aasjournal}

\end{document}